\newcounter{myctr}
\def\myitem{\refstepcounter{myctr}\bibfont\noindent\ifnum\themyctr>9\else\phantom{0}\fi\hangindent17pt\themyctr.\enskip}

\documentclass{ws-ijqi}
\usepackage{tikz}
\usetikzlibrary{quantikz2}
\usepackage{hyperref}
\usepackage[super,sort,compress]{cite}
\let\captionbox\relax
\usepackage{subcaption}
\DeclarePairedDelimiter\abs{\lvert}{\rvert}%
\usepackage{graphicx}
\begin{document}

\catchline{}{}{}{}{}

\title{COMPARATIVE STUDY OF QUANTUM-CIRCUIT SCALABILITY IN A FINANCIAL PROBLEM}

\author{JAEWOONG HEO}

\address{Orientom
\\
Seoul, 06164, Republic of Korea\\
jwheo@orientom.com}

\author{MOONJOO LEE}

\address{Department of Electrical Engineering, Pohang University of Science and Technology\\
Pohang, 37673, Republic of Korea\\
moonjoo.lee@postech.ac.kr}

\maketitle

\begin{abstract}
Quantum computer is extensively used in solving financial problems. Quantum amplitude estimation, an algorithm that aims to estimate the amplitude of a given quantum state, can be utilized to determine the expectation value of bonds as the logic introduced in \textit{quantum risk analysis}. As the number of the evaluation qubit increases, the more accurate the precise the outcome expectation value is. This augmentation in qubits, however, also leads to a varied escalation in circuit complexity, contingent upon the type of quantum computing device. By analyzing the number of two-qubit gates in the superconducting circuit and ion-trap quantum system, this study examines that the native gates and connectivity nature of the ion-trap system lead to less complicated quantum circuits. Across a range of experiments conducted with one to nineteen qubits, the examination reveals that the ion-trap system exhibits a two to three factor reduction in the number of required two-qubit gates when compared to the superconducting circuit system.
\end{abstract}


\markboth{J. Heo \& M. Lee}
{Comparative Study of Quantum-Circuit Scalability in a Financial Problem}

\section{Introduction}

Quantum computers are computational devices composed of quantum systems, and they are anticipated as an alternative to surpass the limitations of conventional computers by advancing computing power. To realize quantum computing, various theoretical and experimental research have been conducted. In addition, beyond simulating complex quantum systems, various quantum algorithms have been investigated and developed to solve complex problems that are difficult to address classically, by utilizing characteristics of quantum computers such as quantum entanglement and superposition. The outcomes of these efforts are being applied in fields such as quantum chemistry, machine learning, and optimization problems.

Finance is a prime sector where benefits from utilizing quantum computers are expected. Stamatopoulos et al.\cite{stamatopoulos2020option} calculated the prices of various European options through amplitude estimation. Rosenberg et al.\cite{rosenberg2015solving} introduced a method to find the optimal path for profit from currency arbitrage by converting it into a quadratic unconstrained binary optimization (QUBO) form and solved through an adiabatic computation. Or{\'u}s et al.\cite{orus2019forecasting} conducted calculations on quantum computers to model the scenario where the capital among financial institutions has interdependency, calculating the entire financial market's crash following the perturbation of underlying asset prices. Significant studies are also being conducted in portfolio optimization which is a fundamental area of investment. Rebentrost et al.\cite{rebentrost2018quantum} utilized the Harrow-Hassidim-Lloyd (HHL) algorithm to perform inverse matrix calculations in Markowitz's mean-variance portfolio optimization, and Rosenberg et al.\cite{rosenberg2018} studied portfolio optimization by reinterpreting the risk parity optimization problem as a QUBO problem.

Woerner et al.\cite{woerner2019quantum} utilized a quantum version of the Monte Carlo method to perform bond pricing. By comparing the classical Monte Carlo simulation with the quantum version, the potential advantages of the quantum Monte Carlo were discussed. The classical Monte Carlo simulation, a parametric method, uses prior knowledge about the distribution of factors constituting the probability for sampling. When estimating the expectation value, it draws $M$ samples from the probability distribution and takes the weighted normalized sum of these samples to estimate the expectation value. It is considered to yield the most accurate estimates if the distribution assumptions hold, though accurate estimation requires many simulation samples. Similarly, the quantum version of the Monte Carlo also requires a significant number of samples to estimate accurate solutions. Generating more samples for the quantum Monte Carlo method requires increasing the number of qubits, a task that presents significant practical challenges. As more qubits are added, the quantum circuit needs to increase in both width and depth to incorporate these additional qubits into the computation process. However, due to current hardware limitations, there are practical limits to circuit expansion. Even for the same computation, the size of the circuit can vary significantly depending on the type of hardware. The type of quantum computer being used determines the implementable gates, and the connectivity between each qubit also becomes a challenge in creating circuits that function.

The paper begins in the \textbf{Background} section by first defining the financial problem at hand and introducing methods to solve this problem using quantum computers. It then introduces the differences between the two types of quantum computer systems. In the \textbf{Results and Discussion} section, this paper analyzes the increase in circuit size, especially the number of two-qubit gates, due to hardware when translating quantum financial algorithms into the actual quantum circuits.

\section{Background}

\subsection{Financial Concept}

Pricing is a fundamental concept in finance, particularly in establishing the fair value of financial instruments like derivatives and stocks. Bonds, being prominent among these instruments, enable entities such as individuals, corporations, and governments to raise capital by issuing them instead of seeking traditional loans. When individuals or entities purchase bonds, they essentially lend money to the bond issuer, making the issuer a debtor and the bond investor a creditor. Unlike loans, bonds are classified as securities and are traded in the market after issuance. Bonds represent debt securities where the issuer commits to repaying the principal borrowed from investors along with predetermined interest at a specified maturity date. Bonds typically provide regular interest payments until maturity, when the principal is repaid. 

\begin{equation}
P_n = P_0 (1+r)^n.
\label{eq:bond payback}
\end{equation}

Eq. (\ref{eq:bond payback}) depicts the future value of bonds where $n$ represents the number of periods, $P_n$ the future price after $n$ period, $P_0$ the initial price, and $r$ the interest rate. A prime example of such bonds is the Treasury Bill (T-Bill), chosen for its simplicity and emblematic status as a short-term security issued by the U.S. Department of the Treasury. These are short-term government securities with maturities of one year or less, issued in the form of zero-coupon bonds, meaning they do not pay interest. The selection of T-Bills for this discussion is due to their straightforward structure and their representation of the most traditional form of short-term government securities. Despite offering relatively low yields, T-Bills are favoured in the short-term financial market due to their minimal risk of default and ample supply. Zero-coupon bonds like T-Bills are issued at a price below their face value and do not pay interest during their tenure. Instead, they are redeemed at face value upon maturity, effectively compensating investors through capital appreciation. It is important to note that bonds carry a time value of money, meaning that the value of a given sum varies depending on when it is received. This variation is influenced by factors such as the risk-free rate and inflation. When interest rates rise above the rate at which T-Bills were purchased, their market value decreases. This relationship between the interest rate changes $\delta r$ and the bond value can be expressed using the following Eq. (\ref{eq:expected value of bond}), assuming interest rate changes are positive and denoting the probability of no change in interest rates as $p$.

\begin{equation}
V = \frac{(1-p)V_F}{1+r+\delta r} + \frac{pV_F}{1+r}.
\label{eq:expected value of bond}
\end{equation}

The above describes the value of a bond where $V$ represents the value of the bond and $V_F$ the face value of the bond. The provided Eq. (\ref{eq:expected value of bond}) can be interpreted as the single-step version of the binomial tree model. This binomial tree model is commonly used to determine the price of securities and can be extended to various financial instruments such as options. 

\begin{figure}[ht]
\begin{center}
    \begin{tikzpicture}
      \tikzstyle{every node}=[circle, draw, fill=white, 
                              minimum size=2.5em, inner sep=0pt]
    
      \node (A) at (-1,0) {$V_{F}$};
      \node (B) at (3,1) {$V_{high}$};
      \node (C) at (3,-1) {$V_{low}$};
    
      \draw [->] (A) -- (B) node [midway, above, draw=none, fill=none] {$p$};
      \draw [->] (A) -- (C) node [midway, below, draw=none, fill=none] {$1-p$};
    
    \end{tikzpicture}
\end{center}
\caption{Description of the changes of value of T-Bill in the single period binomial model where the face value $V_F$ changes to $V_{high}$ with probability $p$ and $V_{low}$ with probability $1-p$.}
\label{fig:binomial tree}
\end{figure}
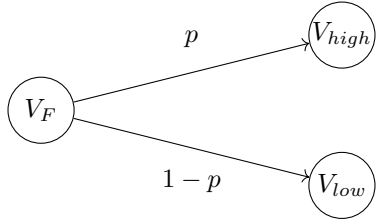

Figure~\ref{fig:binomial tree} represents a T-Bill using the single period binomial tree model. Here, $V_{low}$ represents the scenario where bond value decreases due to rising interest rates, while $V_{high}$ represents the scenario where bond value remains high due to unchanged interest rates. The face value of the bond can either become $V_{high}$ with a probability of $p$ or $V_{low}$ with a probability of $1-p$. Therefore, the value of the bond can be expressed as follows:

\begin{equation}
V = (1-p)V_{low} + pV_{high}.
\label{eq:bond change simplify}
\end{equation}

Eq. (\ref{eq:bond change simplify}) delineates the present value of the bond as a weighted average of potential outcomes. While Eq. (\ref{eq:bond change simplify}) can be solved analytically, Woerner transformed it into a problem of estimating the expectation value and addressed it using quantum Monte Carlo simulation. Bonds like T-Bills serve as underlying assets for other financial products, and such assets can accumulate repeatedly to form complex derivative products. Calculating the fair price of a nonlinear tree constructed from these models can be time-consuming. To address this, various efforts have been made to Monte Carlo simulation with methods such as the delta-gamma approach\cite{glasserman1999importance}, which employs Taylor approximation to simplify and accelerate the computation. Although this combined method can reduce computational complexity by approximating the change in the model's value due to perturbations of underlying probability factors, it may still harbour errors based on the accuracy of the second-order Taylor expansions. Quantum Monte Carlo simulation presents another alternative, offering an algorithm with faster convergence.

\subsection{Expectation Value of Treasury Bill}

The core of Monte Carlo simulation fundamentally involves estimating the expectation output from a given function $f$\cite{montanaro2015quantum}. This process entails generating $M$ independent samples from mathematical model $f$ and computing their average to approximate the expectation output. Following the quantum Monte Carlo integration initiative by Abrams and Williams\cite{abrams1999fast}, advancements have been made from sampling uniform distributions\cite{heinrich2002quantum} to the generalized approaches\cite{brassard2002quantum}. Woerner explored the application of quantum amplitude estimation (QAE) for estimating the expectation value of the simplified T-Bill, executing this on a quantum computer with four to five qubits, thereby illustrating quantum computing's prospective utility in finance.

QAE is a specific quantum algorithm to estimate the amplitude of a certain state within a quantum system through a combination of quantum phase estimation (QPE) followed by quantum amplitude amplification (QAA). QAA is a generalization of Grover's search algorithm which enables access to specific items within an unstructured database composed of multiple objects by amplifying the amplitude of the desired quantum state. The search problem can be modelled as finding $x$ which satisfies $f(x)=1$ when $f:\{0,1,...,M-1\}\rightarrow \{0,1\}$.  In Grover's algorithm, the vectors of the target objects form a subspace in the Hilbert space. If we define a set $G$ of target objects and a set $B$ of non-target objects, the state $\ket{\psi}_n$ after applying the Grover operator $\mathcal{A}$ on the $n$ qubit register can be described as follows.

\begin{equation}
\ket{\psi}_n = \mathcal{A} \ket{0}_{n} = \sqrt{1-a}\ket{\psi_b}_n + \sqrt{a}\ket{\psi_g}_n.
\end{equation}

Here, $a \in [0,1]$ and $\ket{\psi_b}$ and $\ket{\psi_g}$ represent the superposition of eigenvectors corresponding to the set $B$ and $G$ respectively. To make the problem simple, we map $V_{low}=\$0$ and $V_{high}=\$1$ which correspond to the state $\ket{0}$ and $\ket{1}$ so the distribution of expectation value can be mapped with the single objective qubit. Under this simplified single qubit representation, the superposed orthogonal eigenvectors $\ket{\psi_b}_n$ and $\ket{\psi_g}_n$ can be reinterpreted to $\ket{0}$ and $\ket{1}$ respectively.

\begin{equation}
\ket{\psi} = \mathcal{A} \ket{0} = \sqrt{1-a}\ket{0} + \sqrt{a}\ket{1}.
\label{eq:simplified A}
\end{equation}

Eq. (\ref{eq:simplified A}) illustrates that the $\mathcal{A}$ gate encodes the distribution of potential outcomes for $V$ along with their respective probabilities and the objective function by substituting $a$ with probability $p$. This simple operation is analogous to $\mathcal{R_Y}(\theta)$ gate with $\theta=2\sin^{-1}\left({\sqrt{p}}\right)$ as shown below.

\begin{align}
\mathcal{A} \ket{0} 
& = \sqrt{1-p}\ket{0} + \sqrt{p}\ket{1} \notag\\
& = \sqrt{1-\sin^2{\left(\frac{\theta}{2}\right)}}\ket{0} + \sqrt{\sin^2{\left(\frac{\theta}{2}\right)}}\ket{1} \notag\\
& = \cos{\left(\frac{\theta}{2}\right)}\ket{0} + \sin{\left(\frac{\theta}{2}\right)}\ket{1} \notag\\
& = \mathcal{R_Y}(\theta)\ket{0}.
\end{align}

Using QAE, the probability of measuring $\ket{1}$ can be approximated which equals a square of $\sqrt{p}$. This is done using $\mathcal{Q} = -\mathcal{R}_{\psi}\mathcal{R}_{g}$ where $\mathcal{R}_{\psi}$ represents the reflection on $\ket{\psi}$ and $\mathcal{R}_{g}$ the reflection on $\ket{\psi_g}$. Applying this general gate to our simplified state yields:

\begin{align}
\mathcal{Q}\ket{1}
& = -\mathcal{R}_{\psi}(\mathcal{I}-2\ket{1}\bra{1})\ket{1} \notag\\
& =  \mathcal{R}_{\psi}\ket{1}        \notag\\
& =   \left(\mathcal{I}-2\ket{\psi}\bra{\psi}\right)\ket{1}       \notag\\
& =  \ket{1} -2\ket{\psi}\bra{\psi}\ket{1}       \notag\\
& =   \ket{1}-2\left(\cos\left(\frac{\theta}{2}\right)\ket{0}+\sin\left(\frac{\theta}{2}\right)\ket{1}\right)   \sin\left(\frac{\theta}{2}\right)    \notag\\
& =   \left(1-2\sin^2\left(\frac{\theta}{2}\right)\right)\ket{1} -  2\sin\left(\frac{\theta}{2}\right)\cos\left(\frac{\theta}{2}\right)\ket{0}      \notag\\
& =  - \sin \left(\theta\right) \ket{0} + \cos \left(\theta\right) \ket{1}.
\end{align}

Above expansion can be similarly applied to $\ket{0}$, resulting $\mathcal{Q}\ket{0}=\cos \left(\theta\right) \ket{0} + \sin \left(\theta\right) \ket{1}$. Hence, $\mathcal{Q}$ gate which also spans on $\{\ket{0}, \ket{1}\}$ can again be expressed as rotation gate as Eq. (\ref{eq:Qmatrix}).

\begin{equation}
\mathcal{Q} = 
\begin{pmatrix}
\cos\left(\theta\right) & -\sin\left(\theta\right)\\
\sin\left(\theta\right) & \cos\left(\theta\right)
\end{pmatrix} 
= \mathcal{R_Y}(2\theta).
\label{eq:Qmatrix}
\end{equation}

The different geometric powers of the $\mathcal{Q}$ gate on the objective qubit are controlled by $n$ evaluation qubits which are prepared into equal superposition by the set of Hadamard gates $\mathcal{H}$. These $n$ qubits are responsible for the accuracy and resolution of the expectation value approximation. 

\begin{figure}[ht]
\centering
\begin{quantikz}
\lstick{\ket{0}} & \qwbundle{n} & \gate{\mathcal{H}} & \ctrl{1} & \gate{\mathcal{QFT^{\text{-1}}}} &\meter{}
\\
\lstick{\ket{\psi}} & \qw  & \qw       & \gate{\mathcal{Q}} & \qw           &\qw
\end{quantikz}
\caption{Simplified QPE circuit with the state $\mathcal{A}\ket{0}$ prepared on the objective qubit as $\ket{\psi}$ and $n$ bundle of $\ket{0}$ prepared on the evaluation qubits. $\mathcal{Q}$ denotes the controlled phase rotation gate and $\mathcal{H}$ denotes the Hadamard gates acting on each of the evaluation qubits, which later passed on inverse quantum Fourier transform gate $\mathcal{QFT^{\text{-1}}}$. These evaluation qubits were measured in the end by the measurement gate, denoted as the meter symbol in the end.}
\label{fig:qpe circuit}
\end{figure}
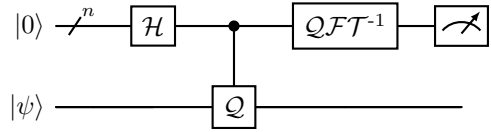

Figure~\ref{fig:qpe circuit} shows the simplified version of QPE with state prepared with $\mathcal{A}$. Here, a set of $\mathcal{H}$ gates are applied to a bundle of evaluation qubits initialized to $\ket{0}$ state. The evaluation qubits control $\mathcal{Q}$ gate on the objective qubit and later pass inverse quantum Fourier transform gate $\mathcal{QFT^{\text{-1}}}$. One thing to note is that the state $\ket{\psi}$ prepared with $\mathcal{A}$ can be expressed as linear combination $\mathcal{Q}$'s eigenvectors of $\ket{\psi_+}$ and $\ket{\psi_-}$. To see how such a circuit can approximate the probability $p$, let's assume that the state $\ket{\psi_+}$ and $\ket{\psi_-}$ can independently be prepared instead of $\ket{\psi}$. Since the eigenvalues of $\mathcal{Q}$ are $\lambda_\pm=\cos\left(\theta\right) \pm i\sin\left(\theta\right) = e^{\pm i\theta}$, the amplitudes of each eigenvectors of $\mathcal{Q}$ can be deduced as below.

\begin{align}
    & \mathcal{Q}\ket{\psi_+} = e^{i\theta}\ket{\psi_+} = e^{2 \pi i \frac{\theta}{2 \pi}}\ket{\psi_+} \notag\\
    & \mathcal{Q}\ket{\psi_-} = e^{-i\theta}\ket{\psi_-} = e^{2 \pi i \frac{2 \pi - \theta}{2 \pi}}\ket{\psi_-}.
\end{align}

If the measurement results of the above states after passing $\mathcal{QFT^{\text{-1}}}$ operation are $x$ and $y$ for $\ket{\psi_+}$ and $\ket{\psi_-}$ respectively, the relationship between measurement results and $\theta$ can be made using double angle identity for the sine function.

\begin{align}
\text{i)} & \quad \text{For } \ket{\psi} = \ket{\psi_+}: \notag\\
& \quad x = 2^{n}\frac{\theta}{2 \pi} \notag\\
& \quad \Rightarrow \theta = \frac{2 \pi x}{2^{n}} \notag\\
& \quad \Rightarrow \sin^2 \left(\theta\right) = \sin^2 \left(\frac{2 \pi x}{2^{n}}\right) \notag\\
\text{ii)} & \quad \text{For } \ket{\psi} = \ket{\psi_-}: \notag\\
& \quad y = 2^{n}\frac{2 \pi - \theta}{2 \pi} \notag\\
& \quad \Rightarrow \theta = 2 \pi \left( 1- \frac{y}{2^{n}} \right) \notag\\
& \quad \Rightarrow \sin^2 \left(\theta\right) = \sin^2 \left(\frac{2 \pi y}{2^{n}}\right).
\end{align}

The above expansions show that the $\sin^2$ values of both answers are the same regardless of the assumed states. Since the original state $\ket{\psi}$ is a linear combination of $\ket{\psi_+}$ and $\ket{\psi_-}$, the measurement result is also $\sin^2 \left(\theta\right) = \sin^2 \left(\frac{2 \pi z}{2^{n}}\right)$ when $z$ is the measurement result. Rearranging the relation with $p=\sin^2\left(\frac{\theta}{2}\right)$ yields $\tilde{p} = \sin^2\left(\frac{\pi z}{M}\right)$ where $z \in \{0,1,...,M-1\}$ and $M=n^2$. The error between the true value of $p$ and the approximation $\tilde{p}$ is as followed with probability of at least $8/\pi^2$:

\begin{equation}
\abs{p-\tilde{p}} \leq \frac{\pi}{M} + \frac{\pi^2}{M^2} \leq \mathcal{O}(M^{-1}).
\end{equation}

The estimation $\tilde{p}$ converges to $p$ in $\mathcal{O}(M^{-1})$, a rate that shows quadratically faster speed than that of classical Monte Carlo simulation where the convergence rate is $\mathcal{O}(M^{-1/2})$. With $\mathcal{A}$ gate and its corresponding $\mathcal{Q}$, the QAE circuit for $p=0.2$ and three evaluation qubits is described below.

\begin{figure}[ht]
    \centering
    \begin{subfigure}{.8\textwidth}
        \centering
        \includegraphics[width=1\textwidth]{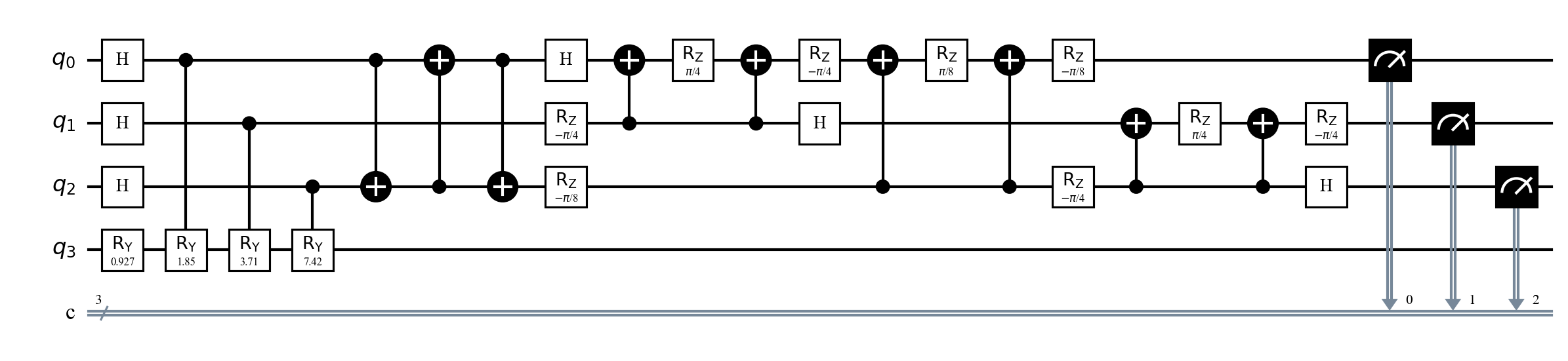}
        \caption{}
        \vspace{0.0cm}
        \label{fig:ideal_raw}
    \end{subfigure}
    \vspace{0.0cm}
    \begin{subfigure}{.8\textwidth}
        \centering
        \includegraphics[width=1\textwidth]{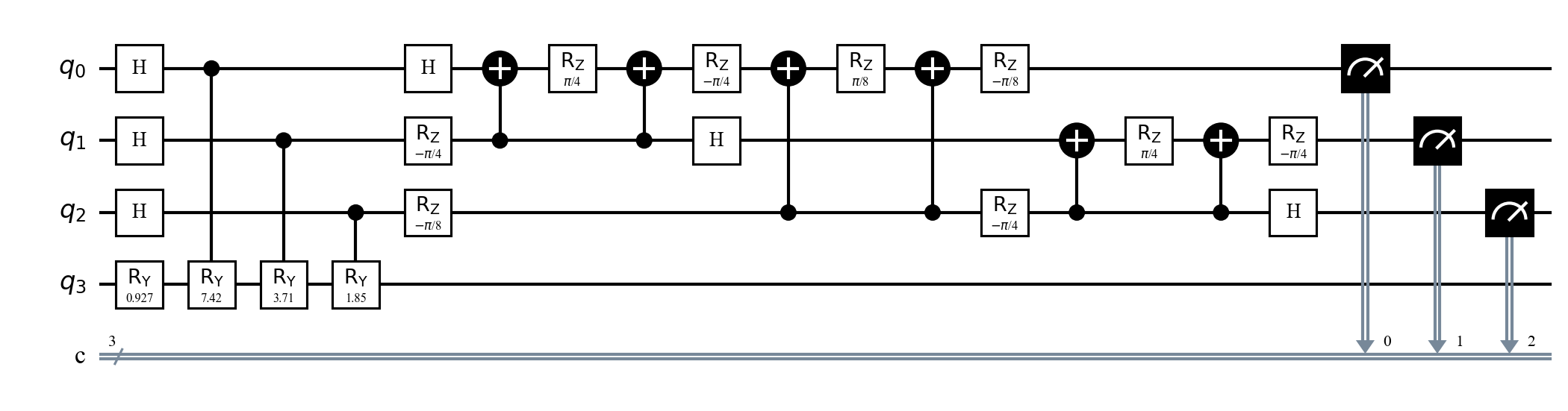}
        \caption{}
        \vspace{0.0cm}
        \label{fig:ideal_opt}
    \end{subfigure}
    \vspace{0.0cm}
\caption{\textit{Qiskit} generated diagrams of the (a) original and (b) optimized QAE circuits for estimating $p=0.2$ with 3 evaluation qubits. The number in the rotation gate represents the rotating angle of the corresponding direction attained through $\theta=2\sin^{-1}\left({\sqrt{p}}\right)$.}
\label{fig:ideal circuit}
\end{figure}

Figure~\ref{fig:ideal circuit} shows the circuit constructed with Hadamard($\mathcal{H}$), Rotation, and CNOT($\mathcal{CX}$) gates. Figure~\ref{fig:ideal_raw} depicts a circuit created automatically using \textit{Qiskit}. This circuit includes an inverse quantum Fourier transform (QFT) that incorporates $\mathcal{SWAP}$ operations, which entail numerous $\mathcal{CX}$ gates thereby demanding significant entanglement operations from quantum hardware. To circumvent this, by altering the sequence of $\mathcal{CR_Y}(\theta)$ gate sampling and eliminating the $\mathcal{SWAP}$ operations, the circuit can be made more lightweight(Figure~\ref{fig:ideal_opt}).

As Figure~\ref{fig:ideal circuit} have three evaluation qubits, they can express $2^3=8$ different bit-strings from $000$ to $111$ which are equivalent to $[\sin^2\left(\frac{\pi \cdot 0}{M}\right),...,\sin^2\left(\frac{\pi \cdot 7}{M}\right)]$. The calculated value represented by the set of bit-strings corresponds to the range from $0$ to $\pi$ of the sine function. Since the number of bit-strings is always an even number, excluding the minimum($000$) and maximum values($100$), two different bit-strings will represent the same value due to the symmetry of the sine function. By combining those pairs, the number of resulting representations forms $\frac{2^3-2}{2}+2=5$ levels of fixed resolution. In this three evaluation qubits case example, the available probability approximations are $[0, 0.146, 0.5, 0.854, 1.0]$.

\begin{figure}[ht]
\centering
    \includegraphics[width=0.5\textwidth]{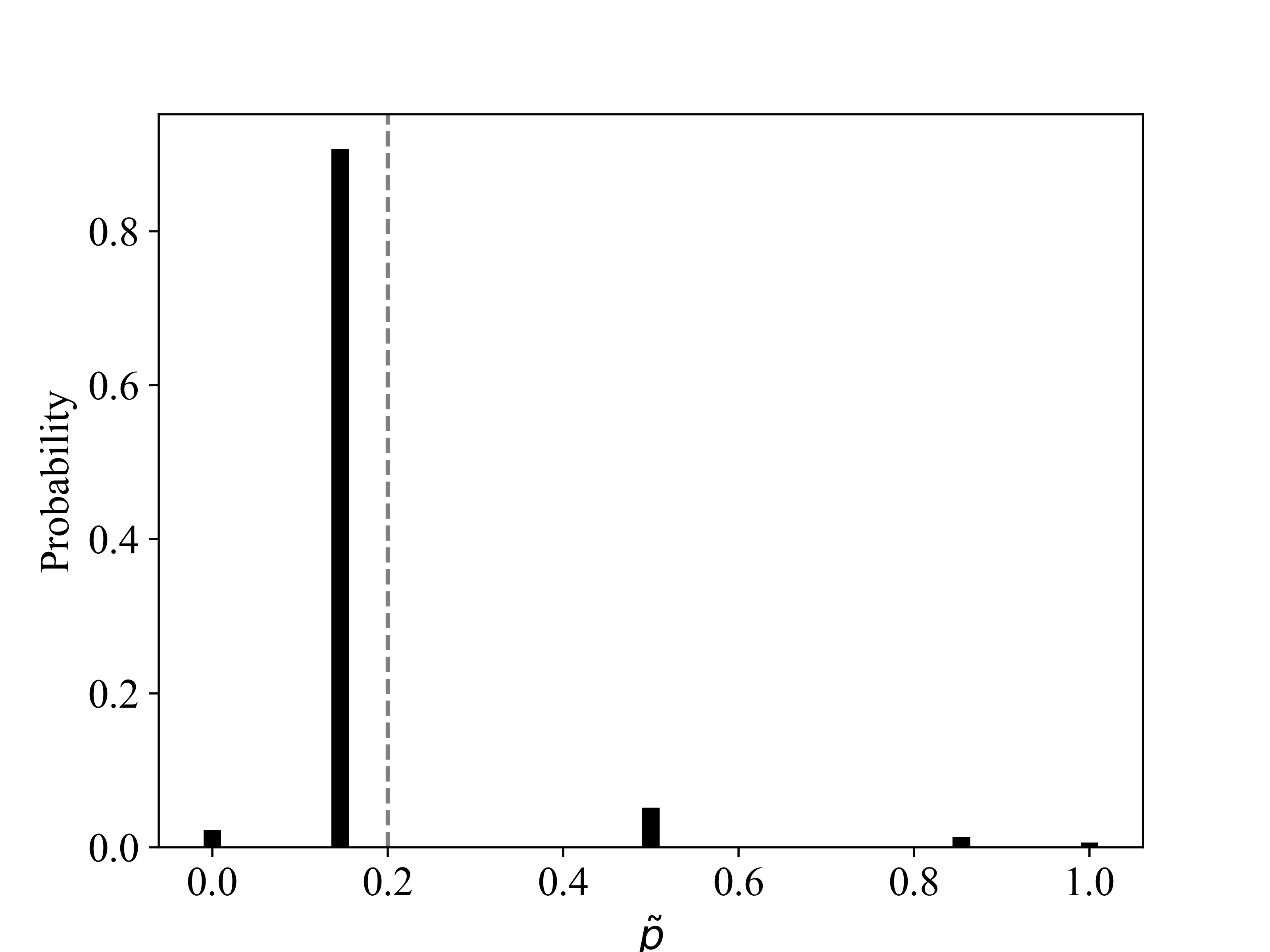}
\caption{Simulation result for $p=0.2$ and with $3$ evaluation qubits, where the x-axis represents the estimated probability $\tilde{p}$ and the y-axis is the normalized number of result population. The true value of $p=0.2$ is represented by the dashed line.}
\label{fig:ideal result}
\end{figure}

Figure~\ref{fig:ideal result} presents the simulation outcomes derived from $10,000$ iterations (shots) using the high-performance quantum simulator by IBM, designed for emulating quantum circuits crafted in the Quantum Assembly Language (QASM). While the true value of $p$ is $0.2$, the simulation cannot estimate the exact value due to limited resolution. The result, as can be seen from Figure~\ref{fig:ideal result}, does not always yield the closest value due to QAE's success probability $8/\pi^2$. 

The final calculation for expectation value can be done classically with the most probable outcome $0.146$, where the analytical calculation is  $V = (1-0.146)\cdot\$0 + 0.146\cdot\$1=\$0.146$. While the steps introduced can be done easily with a classical computer, the key parts of the method introduced here can be used for a more comprehensive circuit which could potentially be used for estimating meaningful financial variables such as option pricing\cite{stamatopoulos2020option}, value at risk\cite{woerner2019quantum}, and credit risk\cite{egger2020credit}.

\subsection{Comparative Analysis of Quantum Systems}

The architecture of a quantum computing system plays a pivotal role in its operational efficiency and scalability. The IBM superconducting quantum system and the IonQ ion-trap system represent two leading approaches, each with distinct advantages and challenges. These platforms differ fundamentally in their physical realization of qubits, the basic units of quantum information, as well as in their operational mechanisms, including how qubits are manipulated and how they interact with each other. This comparative analysis focuses on IBM's superconducting circuit devices and IonQ's ion-trap systems, highlighting their native gates, qubit connectivity, and the implications of these characteristics for quantum computing.

IBM's quantum computers utilize superconducting circuits to create qubits. These circuits operate at extremely low temperatures, close to absolute zero, to maintain superconductivity. Qubits in these systems are typically realized as Josephson junctions, which allow for the creation of superpositions and entanglement, fundamental properties of quantum computing. In terms of connectivity, superconducting qubits are generally arranged in fixed layouts, such as IBM's heavy-hex lattice. The connectivity is determined by the physical placement of the qubits and the resonators that link them. This architecture allows for direct interactions between neighbouring qubits, but interactions between non-neighbouring qubits require the use of $\mathcal{SWAP}$ operations to move quantum information across the chip, potentially leading to increased operation times and error rates. The native gates of IBM's superconducting quantum computers typically include several single-qubit gates and one two-qubit gate, where the current basis gates are $\mathcal{CX}, \mathcal{I}, \mathcal{R_Z}, \mathcal{SX}$, and $\mathcal{X}$\cite{khumalo2022investigation}. These gates form a universal set that can implement any quantum algorithm. The precision and speed of these gates are critical for the performance of the quantum computer, with gate errors and decoherence times being key metrics of system quality.

IonQ's quantum computers use trapped ions as qubits, leveraging the ions' electronic states to encode quantum information. These ions are trapped and isolated in a vacuum chamber using electromagnetic fields. A significant advantage of ion-trap systems is their flexible qubit connectivity. Unlike superconducting qubits, which have fixed neighbours, ions in a trap can be rearranged using electric fields, allowing any qubit to interact directly with any other qubit\cite{chen2023benchmarking}. This all-to-all connectivity reduces the need for $\mathcal{SWAP}$ operations, potentially offering more efficient quantum algorithms. The native gates in ion-trap systems often include single-qubit rotation gates and the Mølmer-Sørensen($\mathcal{MS}(\theta)$) gate, which is a two-qubit entangling gate. Ion-trap quantum computers can precisely control these gates using laser beams, with the ions' motion mediating qubit-qubit interactions. This capability allows for the implementation of high-fidelity operations across the entire qubit register.

\section{Result and Discussion}
In the domain of quantum computing, the implementation of two-qubit gates presents a significantly more complex challenge compared to single-qubit gates. This complexity arises from the intricate manipulation of quantum states required to achieve entanglement, a quintessential quantum phenomenon that underpins the computational advantage of quantum over classical systems. Unlike single-qubit gates, which involve the application of electromagnetic fields to alter the state of an individual qubit, two-qubit operations necessitate a controlled interaction that often involves entangling the qubits. Achieving this level of control demands not only the precise calibration of external fields but also a meticulously engineered qubit environment to facilitate these interactions without introducing significant decoherence or operational errors. The complexity of orchestrating such interactions is substantially greater, given the need to maintain coherence across the qubits involved. Two-qubit gates generally exhibit longer execution times than their single-qubit counterparts, increasing the susceptibility of qubits to decoherence and environmental noise. This extended operation time exacerbates the loss of quantum information, a critical challenge that quantum computing endeavours to overcome. Furthermore, the process of entangling qubits or executing controlled operations amplifies their interaction with the surrounding environment, potentially elevating error rates beyond those associated with single-qubit operations.

\subsection{Two-Qubit Gate Implementations on Quantum Systems}

To measure how many two-qubit gates exist in the circuit in Figure~\ref{fig:ideal_opt}, two types of two-qubit gates should be analysed: $\mathcal{CX}$ and $\mathcal{CR_Y}(\theta)$ gates. $\mathcal{CX}$ can be realized in the IBM system naturally as it is one of the five basis gates of the system. In the IonQ system, the $\mathcal{MS}(\theta)$ gate, which can be applied to multiple ions simultaneously\cite{molmer1999multiparticle}, can be reduced to the two-qubit gate which is equivalent to $\mathcal{R_{XX}}(\theta)= \exp\left( -i\frac{\theta}{2} X\otimes X  \right)$\cite{IBMQuantumDocs}. Using this two-qubit version of $\mathcal{MS}(\theta)$ gate, $\mathcal{CX}$ gate can be constructed as below\cite{maslov2017basic}.

\begin{figure}[ht]
    \centering
    \begin{quantikz}
     & \gate{\mathcal{U}(\frac{\pi}{2},0,0)}& \gate[2]{\mathcal{R_{XX}}(\frac{\pi}{2})}& \gate{\mathcal{U}(\frac{\pi}{2},\frac{\pi}{2},-\pi)} &
    \\
     &                                      &                                & \gate{\mathcal{U}(-\frac{\pi}{2},-\frac{\pi}{2},\frac{\pi}{2})}&
    \end{quantikz}
\caption{$\mathcal{CX}$ gate realization in ion-trap system. \textit{Qiskit} implementation of two-qubit $\mathcal{MS}(\theta)$ gate, the two-qubit $X\otimes X$ interaction gate $\mathcal{R_{XX}}(\theta)$, is placed between the unitary gates $\mathcal{U}(\theta, \phi, \lambda)$ of Eq. (\ref{eq:unitary matrix}).}
\label{fig:CXIonQ}
\end{figure}
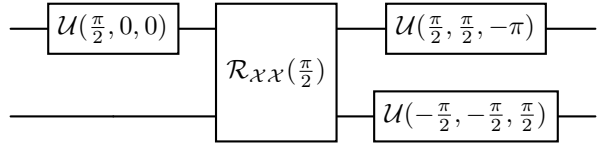

Figure~\ref{fig:CXIonQ} shows one example of how to construct $\mathcal{CX}$ using $\mathcal{R_{XX}}(\theta)$ gate and $\mathcal{U}(\theta, \phi, \lambda)$ gate, where the parameters of unitary $\mathcal{U}$ represents the rotation angle of the Bloch sphere. 

\begin{equation}
\mathcal{U}(\theta, \phi, \lambda) = 
\begin{pmatrix}
\cos \left(\frac{\theta}{2}\right) & -e^{i \lambda}\sin \left(\frac{\theta}{2}\right)\\
e^{i \phi}\sin \left(\frac{\theta}{2}\right) & e^{i \phi + \lambda}\cos \left(\frac{\theta}{2}\right)
\end{pmatrix}.
\label{eq:unitary matrix}
\end{equation}

\begin{figure}[ht]
    \centering
    \begin{subfigure}{.5\textwidth}
    \centering
    \begin{quantikz}
     & \qw                              & \ctrl{1} & \qw                               & \ctrl{1} & \qw
    \\
     & \gate{\mathcal{U}(\frac{\theta}{2},0,0)} & \targ{}  & \gate{\mathcal{U}(\frac{-\theta}{2},0,0)} & \targ{}  & \qw
    \end{quantikz}
    \caption{}
    \label{fig:RyIBM}
    \end{subfigure}
    % \hspace{1em}
    \begin{subfigure}{.5\textwidth}
    \centering
    \begin{quantikz}
     & \gate{\mathcal{U}(\frac{\pi}{2},0,\pi)}      & \gate[2]{\mathcal{R_{XX}}(\frac{-\theta}{2})}& \gate{\mathcal{U}(\frac{\pi}{2},0,\pi)} &
    \\
     & \gate{\mathcal{U}(\frac{3\pi}{2},\frac{\pi}{2},-\pi)}&                                      & \gate{\mathcal{U}(\frac{\pi}{2},0,\frac{\pi}{2})}&
    \end{quantikz}
    \caption{}
    \label{fig:RyIonQ}
    \end{subfigure}
\caption{$\mathcal{CR_Y}(\theta)$ gate realizations on 2 different quantum systems: (a) represents the superconducting circuit system implementation with $\mathcal{U}(\theta, \phi, \lambda)$ gate and $\mathcal{CX}$ gate appear in alternating sequence; (b) represents the ion-trap system implementation with a parametric $X\otimes X$ interaction gate $\mathcal{R_{XX}(\theta)}$ in between the $\mathcal{U}(\theta, \phi, \lambda)$ gates.}
\label{fig:RyBasisGate}
\end{figure}
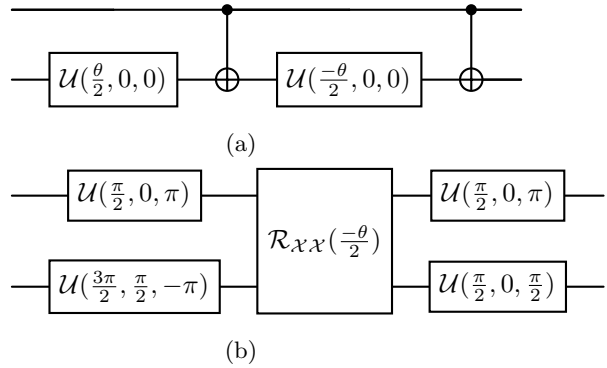

Since the basis gates of IonQ include $\mathcal{R_{X}}(\theta)$, $\mathcal{R_{Y}}(\theta)$, and $\mathcal{R_{Z}}(\theta)$ gates which comprise single qubit universal set, $\mathcal{U}(\theta, \phi, \lambda)$ can be realized by applying appropriate rotation. The same analogy can be applied to the decomposition of $\mathcal{CR_Y}(\theta)$ gate as in Figure~\ref{fig:RyIonQ}.

Figure~\ref{fig:RyBasisGate} shows how the $\mathcal{CR_Y}(\theta)$ gate is decomposed on two different systems. Unlike Figure~\ref{fig:RyIonQ} where the $\mathcal{CR_Y}(\theta)$ gate is decomposed with only a single two-qubit gate, the decomposition of $\mathcal{CR_Y}(\theta)$ gate in IBM system require two $\mathcal{CX}$ gates since the $\theta$ rotation on y-axis when control qubit is in $\ket{1}$ is realized by composition of $\mathcal{R_Y}(\theta/2)$ gate with subsequent $\mathcal{X}\mathcal{R_Y}(-\theta/2)\mathcal{X}=\mathcal{R_Y}(\theta/2)$. Following these two-qubit gates decomposition, the number of two-qubit gates for Figure~\ref{fig:ideal_opt} can be represented as Eq. (\ref{eq:num of 2 qubit ideals}):

\begin{align}
\text{i)}   & \quad \text{IBM}: \notag\\
            & \quad 2\cdot\#\mathcal{CR_Y}(\theta)+1\cdot\#\mathcal{CX}=2\cdot3+1\cdot6=12\notag\\
\text{ii)}  & \quad \text{IonQ}: \notag\\
            & \quad 1\cdot\#\mathcal{CR_Y}(\theta)+1\cdot\#\mathcal{CX}=1\cdot3+1\cdot6=9.
\label{eq:num of 2 qubit ideals}
\end{align}

\subsection{Number of Two-Qubit Gates Varying by Quantum Devices}

\begin{figure}[ht]
    \centering
    \begin{subfigure}{.155\textwidth}
        \centering
        \includegraphics[width=1\textwidth]{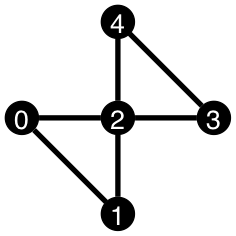}
        \caption{}
        \label{fig:FakeYorktown}
    \end{subfigure}
    \hspace{2.5em}
    \begin{subfigure}{.2\textwidth}
        \centering
        \includegraphics[width=1\textwidth]{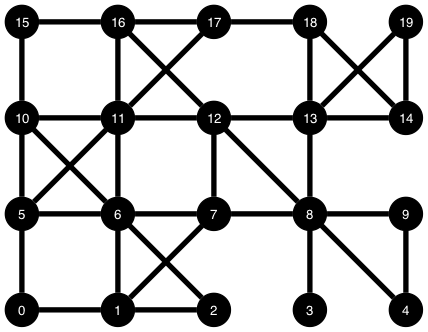}
        \caption{}
        \label{fig:FakeTokyo}
    \end{subfigure}
    \hspace{2.5em}
    \begin{subfigure}{.36\textwidth}
        \centering
        \includegraphics[width=1\textwidth]{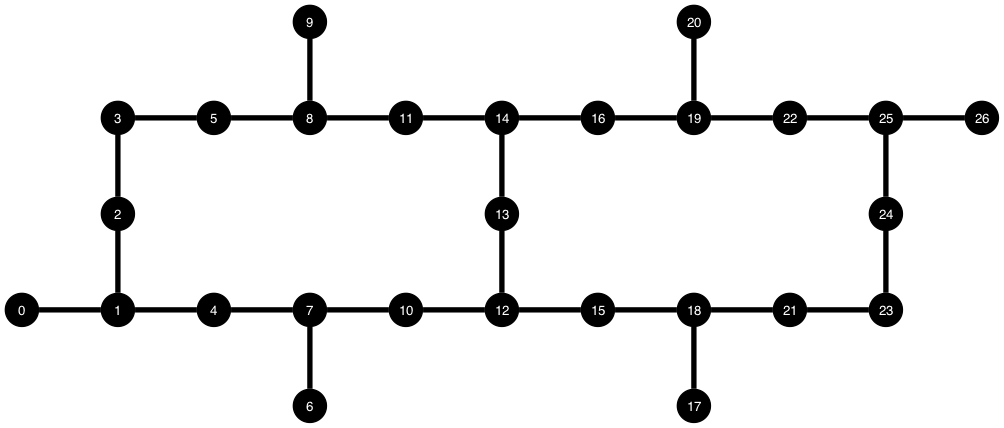}
        \caption{}
        \label{fig:FakeCairo}
    \end{subfigure}
\caption{Different qubit topology in IBM devices: (a) Yorktown has 5 qubits linked in a bow-tie structure; (b) Tokyo has 20 qubits linked in a lattice structure; (c) Cairo has 27 qubits. Each topology representations are marked with indices.}
\label{fig:qubit topology}
\end{figure}

These differences provide advantages for IonQ over IBM, but the differences in two-qubit gates become even more pronounced when considering the actual topology. As mentioned in section 2.3, better connectivity within qubits requires less $\mathcal{SWAP}$ operation for qubit entangling. Since IonQ, as it is built on an ion-trap system, has all-to-all connectivity, the system does not require an additional $\mathcal{SWAP}$ operation for two-qubit entangling. The IBM system, however, has fixed topology varied upon processor architecture as shown in Figure~\ref{fig:qubit topology}.

Figure~\ref{fig:qubit topology} illustrates the qubit topology of three types of IBM quantum processors. Figure~\ref{fig:FakeYorktown} presents the IBM Yorktown with a bow-tie structure, comprising 5 qubits. Figure~\ref{fig:FakeTokyo} depicts the IBM Tokyo which possesses 20 qubits arranged in a 5$\*$4 lattice structure. Figure~\ref{fig:FakeCairo} shows the IBM Cairo, which includes 27 qubits and belongs to the Falcon r5.11 family in IBM's quantum processor classification\cite{IBMQuantumDocs}. The IBM Yorktown and Tokyo, introduced in the \textit{Quantum Risk Analysis} as examples of quantum devices, are currently not available on the IBM quantum cloud, whereas the IBM Cairo is accessible through the service currently.

\subsubsection{All-to-All IBM Device}

In this study, the concept of an 'IBM ideal' is introduced as a hypothetical quantum computing model designed to simulate an all-to-all qubit connectivity scheme, not limited by the physical topological constraint present in existing IBM quantum devices introduced in Figure~\ref{fig:qubit topology}. Unlike these real-world systems, which necessitate qubit swapping due to their specific topological configurations, the 'IBM ideal' assumes a perfect connectivity scenario. This model allows for a more streamlined execution of quantum circuits, requiring only 12 $\mathcal{CX}$ gates as stated in equation~\ref{eq:num of 2 qubit ideals}. The 'IBM ideal' serves as a theoretical benchmark to explore the upper bounds of quantum computing efficiency, free from the topological issues that currently hamper quantum processors. For $n$ evaluation qubits, the number of $\mathcal{CR_Y}(\theta)$ gates is $n$ and the number of $\mathcal{CX}$ gates for $\mathcal{QFT^{\text{-1}}}$ is $n\cdot(n-1)$. Hence, this ideal circuit with $n$ evaluation qubits will require $2*n + n\cdot(n-1)=n^2+n$ two-qubit gates.

\subsubsection{IBM Yorktown and Tokyo}

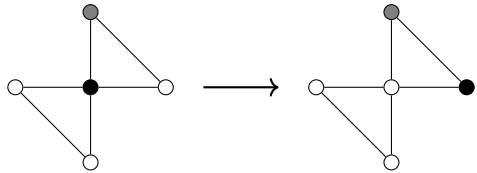
\begin{figure}[ht]
\begin{center}
\begin{tikzpicture}
  \begin{scope}[shift={(0,0)}]
    \node (center) at (0,0) [circle,draw,fill=black,inner sep=2pt]{};
    \node (v1) at (1,0) [circle,draw,fill=white,inner sep=2pt]{};
    \node (v2) at (-1,0) [circle,draw,fill=white,inner sep=2pt]{};
    \node (v3) at (0,1) [circle,draw,fill=gray,inner sep=2pt]{};
    \node (v4) at (0,-1) [circle,draw,fill=white,inner sep=2pt]{};
    
    \draw (center) -- (v1);
    \draw (center) -- (v2);
    \draw (center) -- (v3);
    \draw (center) -- (v4);
    \draw (v1) -- (v3);
    \draw (v2) -- (v4);
  \end{scope}
  
  \draw[->, thick] (1.5,0) -- (2.5,0);
  
  \begin{scope}[shift={(4,0)}]
    \node (center) at (0,0) [circle,draw,fill=white,inner sep=2pt]{};
    \node (v1) at (1,0) [circle,draw,fill=black,inner sep=2pt]{};
    \node (v2) at (-1,0) [circle,draw,fill=white,inner sep=2pt]{};
    \node (v3) at (0,1) [circle,draw,fill=gray,inner sep=2pt]{};
    \node (v4) at (0,-1) [circle,draw,fill=white,inner sep=2pt]{};
    
    \draw (center) -- (v1);
    \draw (center) -- (v2);
    \draw (center) -- (v3);
    \draw (center) -- (v4);
    \draw (v1) -- (v3);
    \draw (v2) -- (v4);
  \end{scope}
\end{tikzpicture}
\end{center}
\caption{Change of qubit placement for Figure~\ref{fig:ideal_opt} in the IBM Yorktown. The left-hand side represents initial qubit placements, where the objective qubit is placed in the middle(black circle). This placement changes to the right-hand side where the objective qubit swaps its position with a evaluation qubit on its right-hand side.}
\label{fig:yorktown change}
\end{figure}

When applying an actual backend, the number of $\mathcal{CX}$ gates can increase depending on the processor's topology. In the case of IBM Yorktown, the qubits 4 or 3 are not connected to the qubits 0 or 1. If the objective qubit was initially assigned to qubit 2, it would be connected to all evaluation qubits, and thus, no $\mathcal{SWAP}$ operations would be needed during the sampling process where three $\mathcal{CR_Y}(\theta)$ gates are applied. However, the subsequent inverse QFT process requires entanglement between the sampling qubits. To achieve this, one of the evaluation qubits needs to move to position 2, meaning that a $\mathcal{SWAP}$ operation between one evaluation qubit and the objective qubit is necessary. Since a $\mathcal{SWAP}$ operation consists of three $\mathcal{CX}$ gates, this additional qubit operation requires a total of 15 entanglement operations for the IBM Yorktown backend. A specific example of such qubit movement is depicted in Figure~\ref{fig:yorktown change}.

The left-hand side of Figure~\ref{fig:yorktown change} shows the initial qubit placement where the objective qubit, depicted as the black circle in the middle, is connected to all other evaluation qubits which are depicted as the outer blank circles. Here, the grey circle represent unused qubits as this example contains only four qubits operations in total. In this stage, all of the $\mathcal{CR_Y}(\theta)$ between the objective qubits and evaluation qubits can be achieved. The $\mathcal{SWAP}$ operation performs a swapping operation between the objective qubit and the evaluation qubit on the rightmost side, resulting in the right-hand side of Figure~\ref{fig:yorktown change} where $\mathcal{CX}$ gate for $\mathcal{QFT^{\text{-1}}}$ operation are taking place.

\begin{figure}[ht]
\centering
    \includegraphics[width=0.95\textwidth]{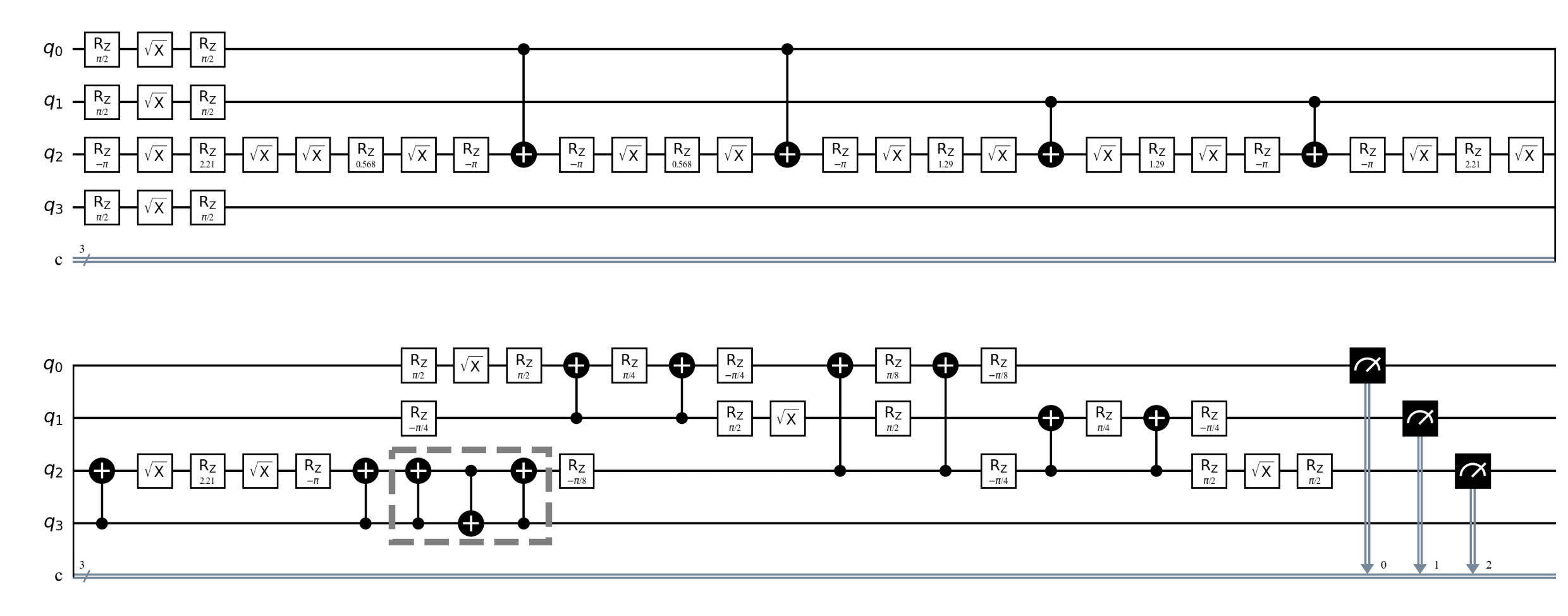}
\caption{Circuit representation of Figure~\ref{fig:ideal_opt} in the IBM Yorktown processor generated with \textit{Qiskit transpile} function. The dashed box indicates the $\mathcal{SWAP}$ operation, composed of three alternating $\mathcal{CX}$ gates, changing the qubit positions as depicted in Figure~\ref{fig:yorktown change}.}
\label{fig:yorktown circuit}
\end{figure}

Figure~\ref{fig:yorktown circuit} shows the circuit representation of the Figure~\ref{fig:ideal_opt} with the scenario from Figure~\ref{fig:yorktown change} for IBM Yorktown device generated with \textit{transpile} function in \textit{Qiskit}. The \textit{transpile} function takes a circuit and backend quantum processor as inputs and generates the actual operation of the corresponding circuit from the backend processor with the gates decomposed to the native gates. As stated, the circuit in IBM Yorktown contains a $\mathcal{SWAP}$ operation at the dashed box which was decomposed into three consecutive $\mathcal{CX}$ gates where the middle one was placed upside down. With an additional $\mathcal{SWAP}$ operation, the total number of two-qubit gates is $12+3=15$. 

In the case of IBM Tokyo, as depicted in Figure~\ref{fig:FakeTokyo}, there is not full connectivity overall, but "partial full connectivity" is observed among four qubits (10, 11, 5, 6). Therefore, if the evaluation qubits and objective qubits are assigned to these positions, no additional $\mathcal{SWAP}$ operations for entangling are needed, resulting in the circuit with only 12 $\mathcal{CX}$ gates which are identical to the ideal case. Yet, similar to the IBM Yorktown case, one additional $\mathcal{SWAP}$ operation is required for four evaluation qubits, bringing the total to 23 $\mathcal{CX}$ gates.

\subsubsection{IonQ}

The trends of adding the $\mathcal{SWAP}$ operation described earlier have an increasingly significant impact as the number of evaluation qubits grows. This is in contrast with the case of the IonQ system where the number of $\mathcal{MS}(\theta)$ gate for $n$ evaluation qubits is $n+n\cdot(n-1)=n^2$, where the former $n$ comes from the number of $\mathcal{CR_Y}(\theta)$ gates and the latter $n\cdot(n-1)$ from the number of $\mathcal{CX}$ gates in the $\mathcal{QFT^{\text{-1}}}$ operator, both of which can be made with single $\mathcal{MS}(\theta)$ gate.

In summary, the exploration of quantum circuit efficiency across various quantum computing devices highlights the significance of device-specific topology on operational complexity. The hypothetical ideal IBM machine, with its all-to-all connectivity, exemplifies an optimal scenario requiring a small number of $\mathcal{CX}$ gates, thereby setting a benchmark for circuit complexity for superconducting circuit systems. In contrast, the example demonstration of IBM Yorktown exhibits the impact of physical qubit connectivity limitation, necessitating additional $\mathcal{SWAP}$ operations and thus increasing the total count of two-qubit gates, which apply similarly to IBM Tokyo and Cairo. Ion-trap system, distinct from the IBM real devices, presents a contrasting approach where the $\mathcal{MS}(\theta)$ gates are directly linked to any qubits due to its all-to-all connectivity, thereby avoiding the compounding complexity introduced by $\mathcal{SWAP}$ operations. This comparative analysis underscores the critical role of device architecture in quantum circuit design.

\subsection{Comparing the Increase in Two-Qubit Gates as the Number of Qubits Grows}

The IBM system possesses different topologies for each quantum processor, necessitating a varying number of $\mathcal{SWAP}$ operations; hence, the increase cannot be described analytically. Additionally, finding the optimal arrangement within a limited number of qubits is challenging, making the required number of $\mathcal{SWAP}$ operations dependent on the initial qubit setting.

\begin{figure}[ht]
  \centering
  \includegraphics[width=0.9\textwidth]{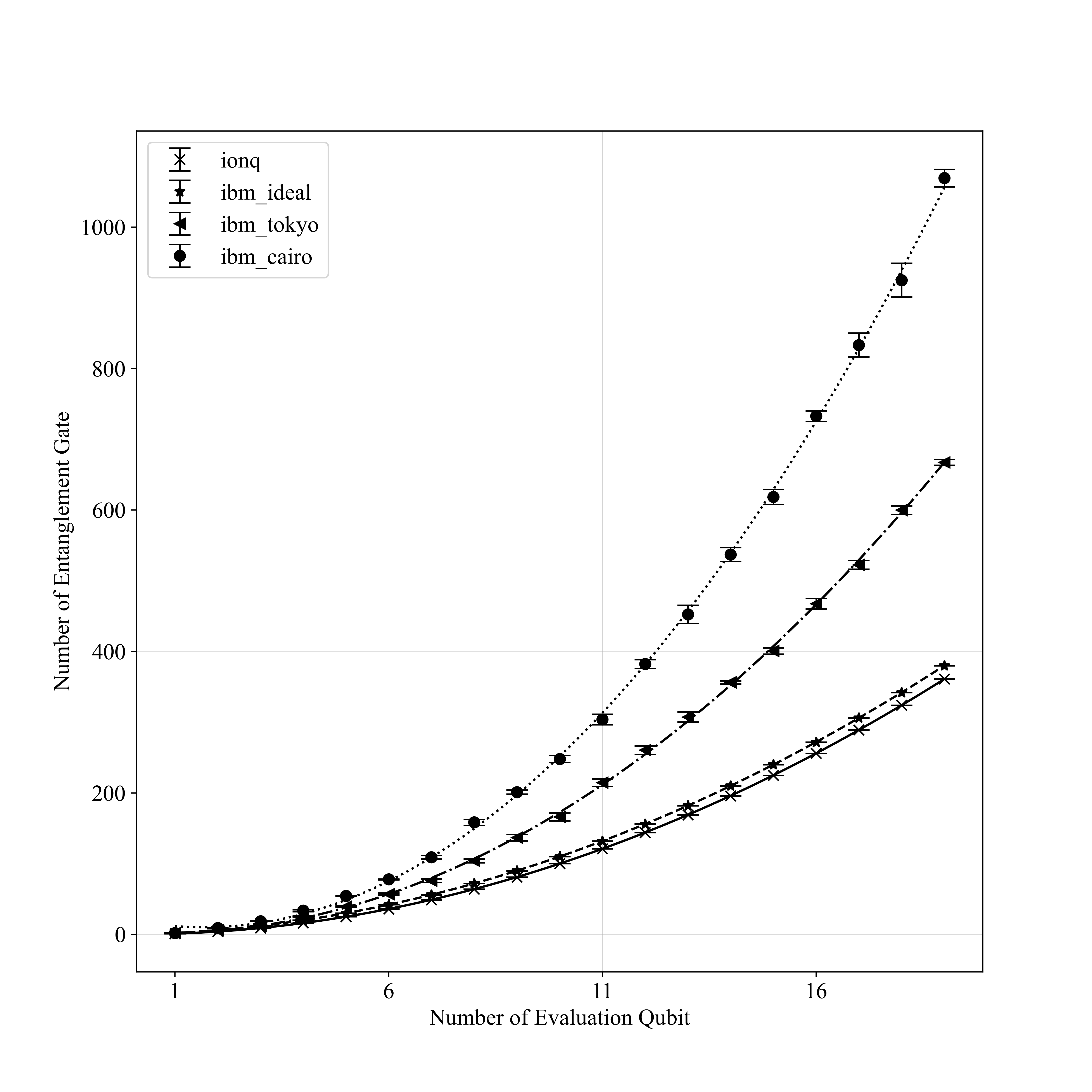}
  \caption{Two-qubit gate count versus evaluation qubit graph, illustrating data points and their polynomial fits for different quantum devices. Each device is represented by a unique marker and line style: IonQ with crosses and solid lines, ideal IBM with stars and dashed lines, IBM Tokyo with triangles and dash-dot lines, and IBM Cairo with circles and dotted lines. The bars associated with each data point indicate the standard error of the mean, calculated as $\frac{s}{\sqrt{n}}$, where $s$ is the standard deviation of the sample, and $n$ is the sample size. This graphical representation allows for the comparison of quantum device performance in terms of two-qubit gate efficiency, as well as the precision of the collected data.}
  \label{fig:gate vs qubit}
\end{figure}

Figure~\ref{fig:gate vs qubit} illustrates, through the data(circle) gathered from counting the number of two-qubit gates in each number of qubits and device type scenario and their corresponding polynomial fit(dotted line), how the number of two-qubit gates increases with the number of evaluation qubits in the four devices. For the simulation, IBM Yorktown was omitted as the device only has five qubits in total. The simulation was conducted up to 19 qubits to see the trend of the number of two-qubit gates growing. The simulation data was attained through \textit{Qiskit} simulation, counting the number of two-qubit gates followed by transpiling each circuit with listed backends. The same circuits were executed multiple times, especially for the IBM real machine analysis, due to the substantial variation in the number of required $\mathcal{SWAP}$ operations depending on the initial qubit placement. The variations of the number of two-qubit gates are shown as the uncertainty bar of each data point, where the deviation of the number can be seen from the data from the IBM real machine for the simulation with a relatively large number of qubits. The second-order polynomial fitting of IonQ and IBM ideal device yields $n^2$ and $n^2+n$ respectively as expected. For IBM Tokyo and IBM Cairo, the polynomials were fitted as $2.1n^2-4.7n+6.8$ and $3.5n^2-12.8n+22.3$ respectively with the same R-squared for both as $0.9988$. This finding suggests that, although the leading order of polynomial fitting remains consistent across the devices, employing an ion-trap or an ideal superconducting circuit system could result in a reduction factor ranging from two to three.

Through the parameters of the polynomials, the efficiency of circuits on various quantum hardware types can be understood. Comprehending how the polynomials are constituted in IBM Tokyo and Cario, however, presents a significant challenge because of the variation of qubit placements during the construction of the circuit in the real hardware. Furthermore, identifying the minimum number of $\mathcal{SWAP}$ operations within a given qubit placement, also known as the qubit routing problem, is NP-hard combinatorial optimization problem\cite{ito2023algorithmic}. SABRE method, which could be applied by altering the optimization level in \textit{Qiskit}, can be a solution for efficient $\mathcal{CX}$ gates insertion which exerts graph search mechanism to search the minimum number of $\mathcal{SWAP}$ operations between two qubits\cite{li2019tackling}. This $\mathcal{SWAP}$ operation mapper, however, is stochastic and still cannot assure the minimum number of $\mathcal{SWAP}$ operations for the circuit.

Similar comparison studies of the two quantum systems have been conducted in the literature. Linke et al. experimented with gates(Margolus and Toffoli) and algorithms(Bernstein-Vazirani and Hidden Shift) on both systems, comparing the success probability and explaining the discrepancy by qubit connectivity, gate fidelity, and error propagation\cite{linke2017experimental}. Lubinski et al. measured the performance benchmark of various quantum hardware by executing various quantum applications\cite{lubinski2023application}. By varying the width and depth of the algorithm on different quantum platforms, the analysis compared the performance across Rigetti, IBM, Quantinuum, and IonQ. Blinov et al. analyse the effect of the two-qubit operation on IBM and IonQ on the measurement result where one of the results about the Bernstein-Vazirani algorithm tells the IonQ shows better fidelity when compared to the IBM Melbourne processor\cite{blinov2021comparison}. While the literature provided explores the difference between the two systems in various aspects, this study only concerns the number of two-qubit gates by scaling the qubit system which accounts for a more precise finance solution, hence focusing more on the number of $\mathcal{SWAP}$ operations. Similar research focusing on the number of two-qubit gates can be extended to the analysis of more a complex bond, which would contain probability distribution loading or non-single interest payment and therefore a more complicated quantum circuit.

\section{Conclusion}
In this paper, the evolution of quantum circuits within two distinct systems (superconducting circuits and ion-trap) is investigated through the T-Bill expectation value problem. A better expectation value necessitates a greater number of evaluation qubits, which in turn requires increased entanglement for the implementation of the quantum amplitude estimation algorithm. Theoretically implemented circuits perform identically in terms of the number of two-qubit gates on the ion-trap system; however, an inconsistent increase in the number of two-qubit gates is required on the superconducting circuits system. This discrepancy is attributed to the fixed topology of superconducting circuits quantum processor, as opposed to ion-trap's all-to-all connectivity, making qubit interaction more challenging and necessitating $\mathcal{SWAP}$ operations for qubit connection. The exact number of required $\mathcal{SWAP}$ operations, being an NP-hard problem, remains unpredictable. Therefore, the circuit size is incrementally enlarged and the number of two-qubit gates is measured using the \textit{transpile} function in \textit{Qiskit}. Data obtained and subsequent fitting into a polynomial graph reveal a difference of a factor of two to three in the highest-order coefficient between ion-trap and superconducting circuits systems, suggesting that the increase in evaluation qubits escalates this disparity, potentially impacting the efficiency of quantum circuits.

\end{document}